\documentclass[preprint,showpacs,titlepage,aps,prd,tightenlines,amsmath,byrevtex,nofootinbib]{revtex4}
\usepackage{graphicx}
\newcommand{\eV}{\text{eV}}

\begin{document}

\preprint{KEK-TH-1033}
\preprint{hep-ph/0508090}

\title{
A formula for the sensitivity to $\sin^2{2\theta_{13}}$ in reactor experiments
with a spectral analysis
}

\author{H.~Sugiyama}
\email{E-mail: hiroaki_at_post.kek.jp}
\affiliation{Theory Group, KEK, Tsukuba, Ibaraki 305-0801, Japan}

\author{O.~Yasuda}
\email{E-mail: yasuda_at_phys.metro-u.ac.jp}
\affiliation{Department of Physics, Tokyo Metropolitan University,
Hachioji, Tokyo 192-0397, Japan}


\vglue 1.4cm
\begin{abstract}
Using an analytical approach,
the sensitivity to $\sin^2{2\theta_{13}}$ with infinite statistics
in a spectral analysis
is investigated in reactor neutrino oscillation experiments
with one reactor and two identical detectors.
We derive an useful formula
for the sensitivity which depends only two parameters $\sigma_{\text{db}}/\sqrt{n}$
(the uncorrelated bin-to-bin systematic error
over the square-root of the number of bins) and
$\sigma_{\text{dB}}$ (the bin-to-bin correlated detector
specific systematic error).
\end{abstract}

\pacs{14.60.Pq,25.30.Pt,28.41.-i}

\maketitle

\section{Introduction}

 A mixing angle $\theta_{13}$ in the lepton sector has not
been measured yet, and we have only an upper bound
obtained by the CHOOZ reactor experiment~\cite{CHOOZ}.
Recently reactor experiments have attracted much attention again
as a possibility to measure $\theta_{13}$%
~\cite{Kozlov:2001jv,Minakata:2002jv,Huber:2003pm,kaska,
Ardellier:2004ui,DiabloCanyon,Braidwood,Daya,Angra,Anderson:2004pk}.
One of the advantages of the reactor measurement of $\theta_{13}$
is possible resolution of the $\theta_{23}$ ambiguity%
~\cite{Fogli:1996pv,Barenboim:2002nv,Minakata:2002jv}.
For the reactor measurement to resolve the $\theta_{23}$ ambiguity,
however, the error in the reactor measurement of $\theta_{13}$
has to be relatively small.  The error in the measurement of
$\sin^22\theta_{13}$ is almost independent of the central value
of $\sin^22\theta_{13}$~\cite{Minakata:2002jv}, so it is basically the same
as the sensitivity to $\sin^22\theta_{13}$.
To resolve the $\theta_{23}$ ambiguity by the reactor measurement,
therefore, the sensitivity to $\sin^22\theta_{13}$ has to be
reasonably small.
Thus estimation of the sensitivity to $\sin^22\theta_{13}$ in
the reactor measurement is important, and for the purpose of
designing reactor experiments in the
future where the statistical error becomes negligibly small
due to large volume of the detectors, it is useful
to know what gives ``the systematic limit'',
i.e.\ the sensitivity to $\sin^22\theta_{13}$
in the limit of infinite statistics.

Refs.~\cite{Kozlov:2001jv,Minakata:2002jv,Huber:2003pm}
discussed the idea of the near-far detector complex
which improves the sensitivity to $\sin^22\theta_{13}$, and
Ref.~\cite{Sugiyama:2004bv} discussed analytically
what gives the dominant contribution to the systematic limit
on $\sin^22\theta_{13}$ by a rate analysis.
In this paper we discuss the systematic limit
using an analytical approach
which was described in detail in~\cite{Sugiyama:2004bv}.
The advantage of the analytical treatment is that
we can easily see which systematic error gives dominant
contribution to the systematic limit
on $\sin^22\theta_{13}$ in the rate as well as spectrum analysis.

Throughout this paper we discuss the case with
a single reactor and two detectors for simplicity
in the limit of infinite statistics.
Also for simplicity, we assume in this paper that
the near and far detectors are identical and have the
same sizes of systematic errors.
In Sect.~2 we review the results of a rate analysis to this system and
show that the dominant contribution to the sensitivity to
$\sin^22\theta_{13}$ comes from the uncorrelated error
$\sigma_{\text{u}}$ between the
detector.  We also derive lower bound on the sensitivity,
which is basically proportional to $\sigma_{\text{u}}$.
In Sect.~3 we perform a spectrum analysis on the same reactor system
and show the dependence of the contribution from each systematic error
on the sensitivity.
In the appendices we give some details
on how to derive the analytic results used in the main text.

\section{Systematic limit of $\sin^2{2\theta_{13}}$
by a rate analysis and its lower bound~\cite{Sugiyama:2004bv,Yasuda:2004dd}}
\begin{figure}
\vspace*{20mm}
\hspace*{-15mm}
\includegraphics[scale=0.8]{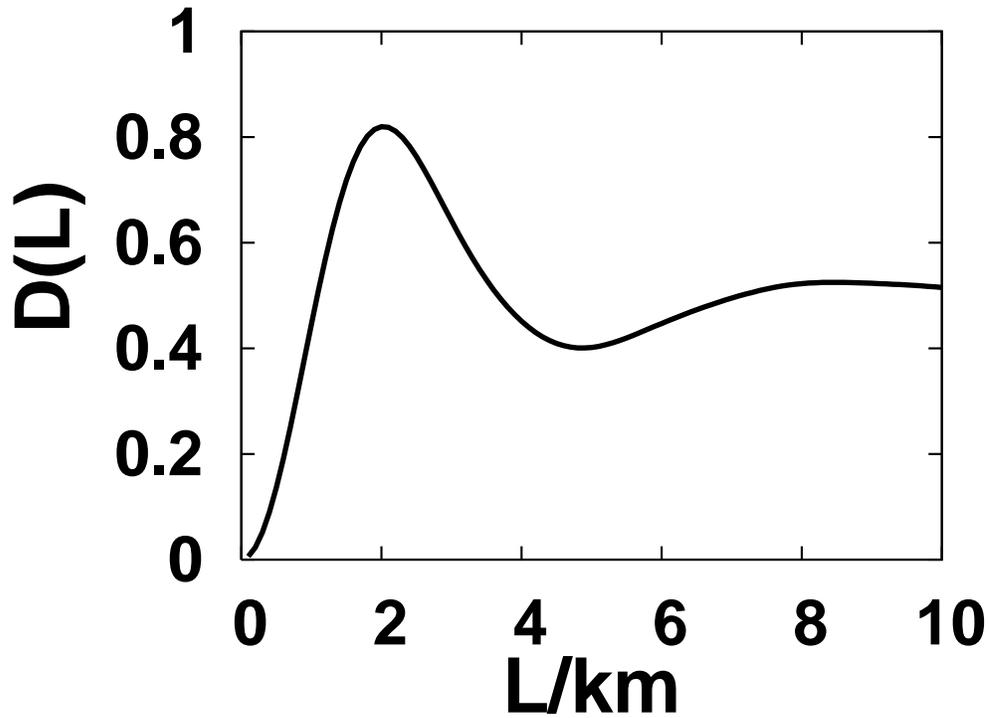}
\vglue 0.5cm
\caption{
$D(L)\equiv \langle
\sin^2\left(\frac{\Delta m^2_{13}L} {4E}\right)\rangle$
as a function of $L$ in the rate analysis.
$D(L)$ has its maximum value 0.82 at
$L$=2.0km, and approaches to its asymptotic value 0.5
as $L\rightarrow\infty$.
$|\Delta m^2_{13}|=2.2\times10^{-3}$eV$^2$
is used as the reference value.
}
\label{fig0}
\end{figure}

Let us first review the sensitivity by a rate analysis in the limit of infinite
statistics.
Let $m^{\text{N}}$ and $m^{\text{F}}$ be the number of events
measured at the near and far detectors, $t^{\text{N}}$ and $t^{\text{F}}$
be the theoretical predictions.
 We must consider the correlation of errors between the detectors.
 $\sigma_{\text{c}}$ and $\sigma_{\text{u}}$
indicate the correlated and uncorrelated systematic errors
in the number of events, respectively.
Then $\chi^2$ is given by~\cite{Sugiyama:2004bv}
\begin{eqnarray}
\hspace*{-5mm}
\displaystyle
\chi^2=
\frac{\left[\left(m^{\text{N}}/t^{\text{N}}-1\right)
+\left(m^{\text{F}}/t^{\text{F}}-1\right)
\right]^2} {4\sigma^2_{\text{c}}
+2\sigma_{\text{u}}^2}
+\frac{\left[\left(m^{\text{N}}/t^{\text{N}}-1\right)
-\left(m^{\text{F}}/t^{\text{F}}-1\right)
\right]^2} {2\sigma^2_{\text{u}}}.
\label{chi2}
\end{eqnarray}
It is straightforward to show that Eq.~(\ref{chi2}) can be rewritten as
\begin{eqnarray}
\chi^2=\sin^42\theta_{13}\left\{
\frac{\left[D(L_{\text{F}})+D(L_{\text{N}})\right]^2} {4\sigma^2_{\text{c}}+2\sigma_{\text{u}}^2}
+\frac{\left[D(L_{\text{F}})-D(L_{\text{N}})\right]^2} {2\sigma^2_{\text{u}}}\right\}
\label{chi3}
\end{eqnarray}
with
\begin{eqnarray}
\displaystyle
D(L)\equiv\left\langle \sin^2\left(\frac{\Delta m^2_{13}L} {4E}
\right) \right\rangle
\equiv
{\displaystyle
\int dE~\epsilon(E)Y(E)\sigma(E)\sin^2\left({\Delta m^2_{13}L \over 4E}
\right) \over \displaystyle
\int dE~\epsilon(E)Y(E)\sigma(E)},
\nonumber
\end{eqnarray}
where $\epsilon(E)$, $Y(E)$, $\sigma(E)$ stand for the detection
efficiency, the neutrino yield at the reactor, and the cross section, respectively.
The numerical value of $D(L)$ is plotted in Fig.~\ref{fig0}
as a function of $L$, where the reference value for
$|\Delta m^2_{13}|$
is 2.2$\times10^{-3}$eV$^2$ which was obtained by
a combined analysis~\cite{Maltoni:2004ei}
of the data of the atmospheric neutrino and
K2K experiments.
Here we assume
$\sigma_{\text{u}}={\cal O}(0.1)\%$ and
$\sigma_{\text{c}}={\cal O}(1)\%$.%
~~$(4\sigma_{\text{c}}^2+2\sigma_{\text{u}}^2)^{-1}$
is much smaller than
$(2\sigma_{\text{u}}^2)^{-1}$
while $D(L_{\text{F}})+D(L_{\text{N}})$ and
$D(L_{\text{F}})-D(L_{\text{N}})$ are generally comparable in magnitude,
so the first term in Eq.~(\ref{chi3})
can be ignored.%
\footnote{
In the region of very large $|\Delta m_{13}^2|$,
the first term dominates the sensitivity because
of $D(L_F)\simeq D(L_N)\simeq 0.5$.}
Hence $\chi^2$ is given approximately by
\begin{eqnarray}
\chi^2\simeq\sin^42\theta_{13}
\frac{\left[D(L_{\text{F}})-D(L_{\text{N}})\right]^2} {2\sigma^2_{\text{u}}}.
\label{chi4}
\end{eqnarray}
The hypothesis of no oscillation is
excluded at the 90\%CL if $\chi^2$ is larger than 2.7, which
corresponds to the value at the 90\%CL for one degree
of freedom.  This implies that the systematic limit
on $\sin^22\theta_{13}$ at the 90\%CL, or the sensitivity
in the limit of infinite statistics, is given by
\begin{eqnarray}
\left(\sin^22\theta_{13}\right)_{\text{limit}}^{\text{sys~only}}
\simeq\sqrt{2.7}\frac{\sqrt{2}\sigma_{\text{u}}} {D(L_{\text{F}})-D(L_{\text{N}})}.
\label{sens0}
\end{eqnarray}
To optimize
$\left(\sin^22\theta_{13}\right)_{\text{limit}}^{\text{sys~only}}$,
therefore, we have to minimize $D(L_{\text{N}})\equiv
\langle \sin^2\left({\Delta m^2_{13}L_{\text{N}} / 4E}
\right)\rangle$ and maximize $D(L_{\text{F}})\equiv
\langle \sin^2\left({\Delta m^2_{13}L_{\text{F}} / 4E}
\right)\rangle$.  Since the possible maximum value of
$D(L_{\text{F}})-D(L_{\text{N}})$ is 0.82,
which is obtained with $L_{\text{F}}=2.0$km and $L_{\text{F}}=0$
for $|\Delta m^2_{13}|=2.2\times 10^{-3}\eV^2$,
the lower bound of
$\left(\sin^22\theta_{13}\right)_{\text{limit}}^{\text{sys~only}}$
in this case is estimated as:
\begin{eqnarray}
\mbox{\rm lower bound of }
\left(\sin^22\theta_{13}\right)_{\text{limit}}^{\text{sys~only}}
\simeq\frac{\sqrt{2.7}\sqrt{2} \sigma_{\text{u}}}{0.82}
=2.8\,\sigma_{\text{u}}.
\label{lbound1}
\end{eqnarray}
Eq.~(\ref{lbound1}) indicates that
the sensitivity is at best $\sin^22\theta_{13}\simeq0.016$
if we adopt the reference value
$\sigma_{\text{u}}=0.6\%$ assumed in Ref.~\cite{Minakata:2002jv}.
To reach this limit of the sensitivity,
we have to set $L_{\text{F}}$
as close to 2.0km as possible, set $L_{\text{N}}$
as close to 0 as possible.
 For the case with finite statistical errors,
$L_{\text{F}}$ which is slightly smaller than 2km is appropriate
because it gives smaller statistical errors.

\section{Systematic limit of $\sin^2{2\theta_{13}}$
by a spectral analysis}
In the previous section we have seen that
the sensitivity to $\sin^2{2\theta_{13}}$ is governed
by the uncorrelated systematic error of the detectors in the rate analysis.
In this section we will examine what happens if
we use spectral information also.

We introduce the systematic errors in almost
the same way as in~\cite{Huber:2003pm}.
$\chi^2$ is given by
\begin{eqnarray}
\hspace*{-20mm}
\displaystyle
\chi^2&=&\min_{\alpha's}\Bigg\{
\displaystyle\sum_{A=N,F}\sum_{i=1}^n
\frac{1} {(t^A_i\sigma^A_i)^2}
\left[m^A_i-t^A_i(1+\alpha+\alpha^A+\alpha_i)
-\alpha_{\text{cal}}^A t^A_iv^A_i
\right]^2\nonumber\\
&+&\displaystyle\sum_{A=N,F}\left[\left(
\frac{\alpha^A} {\sigma_{\text{dB}}}\right)^2
+\left(\frac{\alpha_{\text{cal}}^A} {\sigma_{\text{cal}}}\right)^2\right]
+\displaystyle\sum_{i=1}^n\left(
\frac{\alpha_i} {\sigma_{\text{Db}}}\right)^2
+\left(\frac{\alpha} {\sigma_{\text{DB}}}\right)^2\Bigg\}.
\label{chipull}
\end{eqnarray}
Here, $m^A_i$ is the number of events to be measured
at the near ($A=N$) and far ($A=F$) for the $i$-th energy bin
with the neutrino oscillation,
and $t^A_i$ is the theoretical prediction without the oscillation.
$(\sigma^A_i)^2$ is the uncorrelated error
which consists of the statistical plus uncorrelated bin-to-bin
systematic error:
\begin{eqnarray}
\hspace*{-20mm}
(t^A_i\sigma^A_i)^2
=t^A_i+\left(t^A_i\sigma^A_{\text{db}}\right)^2,
\nonumber
\end{eqnarray}
where $\sigma^A_{\text{db}}$ is the uncorrelated bin-to-bin
systematic error.
For simplicity
we assume that sizes of the bin-to-bin uncorrelated systematic errors
of the detectors and of the flux are independent of the energy.
This may not be the case in practical situations, but
our analytic discussions will be illuminative to see
how each systematic error affects the sensitivity.
$\alpha$ is a variable which corresponds to a
common overall normalization error $\sigma_{\text{DB}}$ for
the number of events.  $\alpha^A~(A=N,F)$ is a variable
which introduces the detector-specific
uncertainties $\sigma_{\text{dB}}$ of the near and far detectors.
$\alpha_i~(i=1,\cdots,n)$ is a variable for
an uncertainty $\sigma_{\text{Db}}$ of the
theoretical prediction for each energy bin which
is uncorrelated between different energy bins.\footnote{
Here we follow the notation for the systematic errors in Ref.~\cite{Minakata:2003wq}.
The first suffix of $\sigma$ stands for the property for the systematic error with respect
to the detectors while the second is with respect
to bins, and capital (small) letter stands for a correlated (uncorrelated) systematic error.
The correspondence for the notation in Ref.~\cite{Huber:2003pm} is as follows:
$\sigma_u=\sigma_{\text{db}}$,
$\sigma_b=\sigma_{\text{dB}}$,
$\sigma_{\text{shape}}=\sigma_{\text{Db}}$,
$\sigma_a=\sigma_{\text{DB}}$.}
$\alpha_{\text{cal}}^A~(A=N,F)$ is a variable which introduces 
an energy calibration uncertainty $\sigma_{\text{cal}}$
and comes in the theoretical prediction in the form
of $(1+\alpha_{\text{cal}}^A)E$ instead of the observed energy $E$.
Thus, the deviation $v^A_i$ (divided by
the expected number of events) from the
theoretical prediction $t^A_i$ due to
this uncertainty can be written as
\begin{eqnarray}
\hspace*{-2mm}
v^A_i=\lim_{\alpha_{\text{cal}}^A\rightarrow0}
\frac{1}{\alpha_{\text{cal}}^A t^A_i}
\left[\frac{N_p T}{4\pi L_A^2}\int_{(1+\alpha_{\text{cal}}^A)E_i}^{(1+\alpha_{\text{cal}}^A)E_{i+1}}dE \epsilon(E) Y(E)\sigma(E)
-t^A_i\right],
\label{v}
\end{eqnarray}
where $N_p$ is the number of target protons in the detector,
$T$ denotes the exposure time,
and $L_A$ is the baseline for the detector $A$.
We have used the definition
\begin{eqnarray}
\hspace*{-20mm}
t^A_i\equiv\frac{N_p T}{4\pi L_A^2}\int_{E_i}^{E_{i+1}}dE
\epsilon(E) Y(E)\sigma(E).
\label{t}
\end{eqnarray}
Our strategy is to
{\it assume no oscillation in the theoretical prediction} $t^A_i$,
to plug the number of events {\it with oscillation} in $m^A_i$,
and to see the sensitivity by looking at the value of
$\chi^2$.  Since $\chi^2$ is quadratical in the variables
$\alpha$, $\alpha^A$, $\alpha_i$, $\alpha_{\text{cal}}^A$, we can minimize
with respect to these variables in Eq.~(\ref{chipull}) exactly.\footnote{
In principle we could take a different convention,
such as multiplying the measured numbers $m^A_i$ by
the uncertainty $(1+\alpha+\alpha^A+\alpha_i)$, etc., and in fact it is
done in some references.
With such a convention, it becomes complicated
to work with an analytical approach because $\chi^2$ is not a Gaussian
with respect to $m^A_i$, which includes oscillation parameters,
after the minimizations of $\alpha$'s.
Note that $\chi^2 = 2.7$ gives a 90\%CL bound for a Gaussian.
 However, the difference between such a
convention and ours affects only the higher orders
in $\sigma^2$'s.
The conclusion on the sensitivity
with such a convention should coincide with ours numerically.}
Hereafter, we take the
limit of infinite statistics, i.e., $1/t^A_i\rightarrow0$
in order to see how systematic errors affect
the sensitivity to $\sin^22\theta_{13}$;
We refer to the sensitivity in this limit as ``the systematic limit''. 
Since we assume
no oscillation for the theoretical prediction $t^A_i$,
the quantity
$v^A_i$ in Eq.~(\ref{v}) is independent of $A(=N,F)$:
\begin{eqnarray}
\hspace*{-20mm}
v^N_i=v^F_i=v_i.
\label{v3}
\end{eqnarray}

After some calculations (see Appendix~\ref{appendix1} for details),
(\ref{chipull}) is rewritten as
\begin{eqnarray}
\hspace*{-30mm}
\chi^2&=&\frac{\sin^42\theta_{13}}{2}
\left\{
\frac{1}{\sigma^2_{\text{db}}/n}\,\cdot\,\frac{1}{n}
\sum_{j=1}^{n-2}
\left[\vec{u}^{{}^{(-)}}_j\cdot(\vec{D}^{F}-\vec{D}^{N})\right]^2\right.
\nonumber\\
&{\ }&+\frac{1}{(\sigma^2_{\text{db}}+\Lambda^{{}^{(-)}}_-)/n}
\,\cdot\,\frac{1}{n}
\left[\vec{u}^{{}^{(-)}}_{n-1}\cdot(\vec{D}^{F}-\vec{D}^{N})\right]^2\nonumber\\
&{\ }&+\frac{1}{(\sigma^2_{\text{db}}+\Lambda^{{}^{(-)}}_+)/n}
\,\cdot\,\frac{1}{n}
\left[\vec{u}^{{}^{(-)}}_n\cdot(\vec{D}^{F}-\vec{D}^{N})\right]^2
\nonumber\\
&{\ }&
+\frac{1}{(\sigma^2_{\text{db}}+2\sigma^2_{\text{Db}})/n}
\,\cdot\,\frac{1}{n}\sum_{j=1}^{n-2}
\left[\vec{u}^{{}^{(+)}}_j\cdot(\vec{D}^{F}+\vec{D}^{N})\right]^2\nonumber\\
&{\ }&+\frac{1}
{(\sigma^2_{\text{db}}+2\sigma^2_{\text{Db}}+\Lambda^{{}^{(+)}}_-)/n}
\,\cdot\,\frac{1}{n}
\left[\vec{u}^{{}^{(+)}}_{n-1}\cdot(\vec{D}^{F}+\vec{D}^{N})\right]^2\nonumber\\
&{\ }&\left.
+\frac{1}{(\sigma^2_{\text{db}}+2\sigma^2_{\text{Db}}+\Lambda^{{}^{(+)}}_+)/n}
\,\cdot\,\frac{1}{n}
\left[\vec{u}^{{}^{(+)}}_n\cdot(\vec{D}^{F}+\vec{D}^{N})\right]^2
\right\},
\label{chi6}
\end{eqnarray}
where we have introduced the variables
\begin{eqnarray}
\hspace*{-20mm}
D^A_i&\equiv&-\frac{1}{\sin^22\theta_{13}}\frac{m^A_i-t^A_i} {t^A_i}
=
{\displaystyle
\int_{E_i}^{E_{i+1}} dE~\epsilon(E)Y(E)\sigma(E)\sin^2\left({\Delta m^2_{13}L_A \over 4E}
\right) \over \displaystyle
\int_{E_i}^{E_{i+1}} dE~\epsilon(E)Y(E)\sigma(E)},
\end{eqnarray}
and $\vec{u}^{{}^{(\pm)}}_j$ are the orthonormal eigenvectors
of some $n\times n$ unitary matrices (see Appendix~\ref{appendix1}).
 $\Lambda^{{}^{(-)}}$ is defined by
$b^{(-)}\equiv\sigma_{\text{dB}}^2$ and $\sigma_{\text{cal}}$
while $\Lambda^{{}^{(+)}}$ is defined by
$b^{(+)}\equiv\sigma_{\text{dB}}^2+\sigma_{\text{DB}}^2$
and $\sigma_{\text{cal}}$;
$\Lambda^{{}^{(-)}}$ ($\Lambda^{{}^{(+)}}$) vanishes
for $b^{(-)}=0$ ($b^{(+)}=0$) or $\sigma_{\text{cal}}=0$,
and then the second (fifth) term in Eq.~(\ref{chi6})
degenerate to the first (fourth) term.
 The first three terms in Eq.~(\ref{chi6}) correspond to
the comparisons of the numbers of events at the two detectors
because these terms does not include
$\vec{D}^{F}+\vec{D}^{N}$ and the correlated errors between the detectors
($\sigma_{\text{Db}}$ and $\sigma_{\text{DB}}$).
 We see that the first and fourth terms in Eq.~(\ref{chi6})
correspond to the ``pure'' spectral analysis
because these terms are free from $\sigma_{\text{cal}}$
and correlated errors among bins
($\sigma_{\text{dB}}$ and $\sigma_{\text{DB}}$);
 Other four terms include the effect of the rate analysis.

\begin{figure}
\vspace*{20mm}
\hspace*{-15mm}
\includegraphics[scale=1.4]{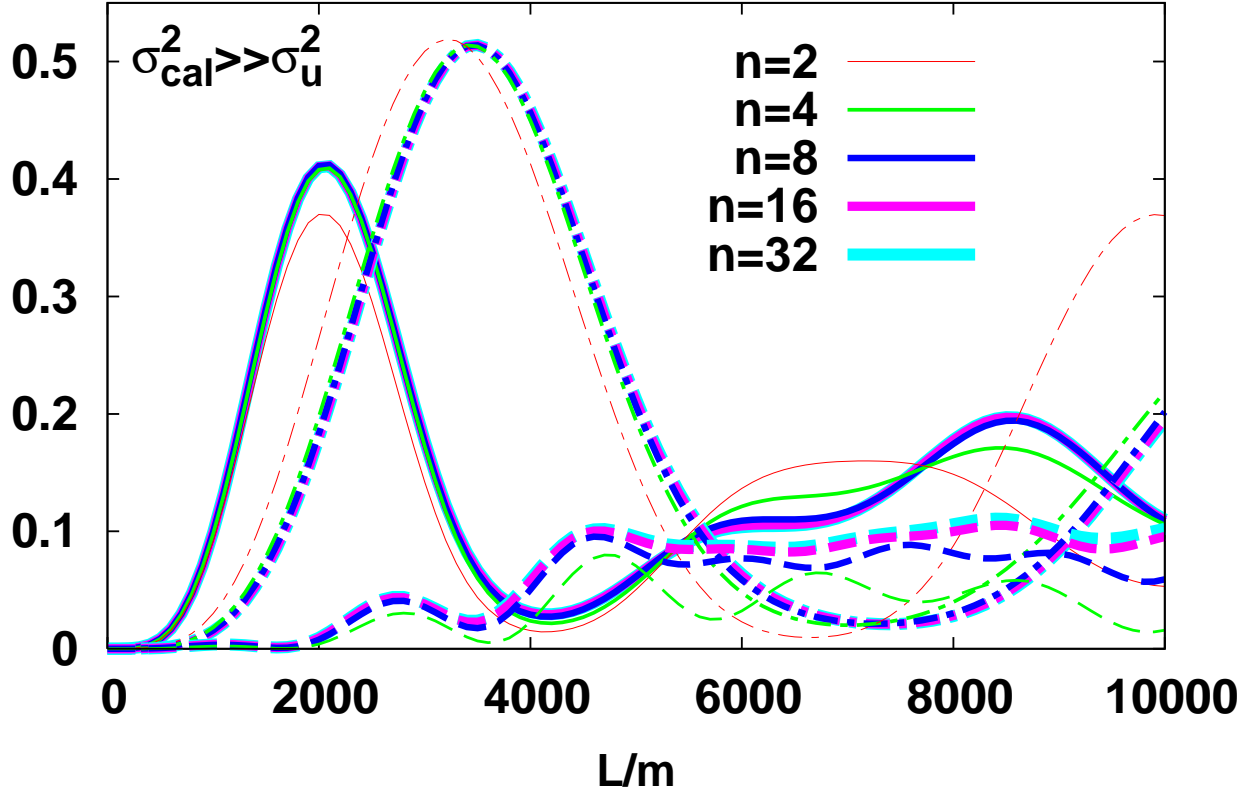}
\caption{
The values of
$\displaystyle\frac{1}{n}
\sum_{j=1}^{n-2}\left(\vec{u}^{{}^{(-)}}_j
\cdot\vec{D}^{F}\right)^2$ (the dash-dotted lines),
$\displaystyle\frac{1}{n}
\left(\vec{u}^{{}^{(-)}}_{n-1}\cdot\vec{D}^{F}\right)^2$
(the thick solid lines) and
$\displaystyle\frac{1}{n}
\left(\vec{u}^{{}^{(-)}}_n\cdot\vec{D}^{F}\right)^2$
(the dashed lines)
for different numbers $n$ of bins
in the limit $\sigma^2_{\text{cal}}\gg\sigma^2_{\text{u}}$,
where $L_F=L$, $L_N=0$.
The latter two are almost saturated for $n\ge4$,
while the first quantity is saturated only for
$n\ge16$.  $|\Delta m^2_{13}|$ is
$2.2\times10^{-3}$eV$^2$.
}
\label{fig3}
\end{figure}

Here we will take the following assumptions for the systematic errors:
\begin{eqnarray}
\sigma_{\text{db}}&\lesssim&\sigma_{\text{u}}={\cal O}(0.1)\%,\nonumber\\
\sigma_{\text{dB}}&\simeq&\sigma_{\text{u}}={\cal O}(0.1)\%,\nonumber\\
\sigma_{\text{Db}}&\lesssim&\sigma_{\text{c}}={\cal O}(1)\%,\nonumber\\
\sigma_{\text{DB}}&\simeq&\sigma_{\text{c}}={\cal O}(1)\%,\nonumber\\
\sigma_{\text{cal}}&=&{\cal O}(0.1)\%.
\label{error}
\end{eqnarray}
Roughly speaking, the errors must satisfy the relations
$\sigma_{\text{u}}^2 \simeq \sigma_{\text{dB}}^2+\sigma_{\text{db}}^2/n$
and $\sigma_{\text{c}}^2 \simeq \sigma_{\text{DB}}^2+\sigma_{\text{Db}}^2/n$,
where $\sigma_{\text{u}}$ and $\sigma_{\text{c}}$ are the errors
appear in the rate analysis (\ref{chi3}).
Among the reactor experiments in the past,
Bugey~\cite{Declais:1994su} seems to be the only
experiment in which the energy spectrum analysis was performed
with the identical detectors.
We have learned~\cite{Stutz} that
$\sigma_{\text{db}}$ in the Bugey experiment was of order of 0.5\%.
We have checked numerically that each factor
$\displaystyle\frac{1}{n}\sum_{j=1}^{n-2}\left[\vec{u}^{{}^{(\pm)}}_j
\cdot(\vec{D}^{F}\pm\vec{D}^{N})\right]^2$,
$\displaystyle\frac{1}{n}\left[\vec{u}^{{}^{(\pm)}}_{n-1}
\cdot(\vec{D}^{F}\pm\vec{D}^{N})\right]^2$,
$\displaystyle\frac{1}{n}\left[\vec{u}^{{}^{(\pm)}}_n
\cdot(\vec{D}^{F}\pm\vec{D}^{N})\right]^2$
is approximately independent of the number $n$ of bins for
$n\gtrsim8$,\footnote{
We found from numerical calculations
that the results with different numbers ($n=2^m, m=3,\cdots,6$) of bins
do not change much.  Fig.~\ref{fig3} gives the dependence
of the first three quantities in Eq.~(\ref{chi6}) on the
number $n$ of bins for $L_F=L$ and $L_N=0$, and it shows
that the behaviors for $n\ge8$ are almost the same.
}
and the maximum value of these factors range from 0.4 to 2 (See Table
\ref{table}).  In most of the analyses in the present paper we take
the number $n$ of bins $n=16$ and the energy interval 2.8MeV$\le
E_\nu\le$7.8MeV.
The optimized baselines $L_F$ and $L_N$ do depend on
the energy interval of the analysis because different
energy interval gives different average of the
neutrino energy.  We adopt the
energy interval 2.8MeV$\le E_\nu\le$7.8MeV because (i) the same
energy interval was used in the analysis of the Bugey experiment,
and (ii) the value of $\Lambda^{{}^{(-)}}_-/n$ becomes
approximately independent of the number $n$ of bins.
(ii) is not necessarily the case, for example, if we take the
energy interval 1.8MeV$\le E_\nu\le$7.8MeV because in this case
the value of $|\vec{v}|$ becomes so large near the threshold energy
$E_\nu\sim1.8$MeV that $\Lambda^{{}^{(-)}}_-/n$
would depend on $n$.
 In fact, we can neglect such an energy region near the threshold
because the region has only small number of events
and does not affect the value of $\chi^2$ so much.

From the assumed values in Eq.~(\ref{error}) it follows that the
last three terms in Eq.~(\ref{chi6})
are negligible because the
coefficients in front of these three factors are much smaller
than those in front of the first three and because
numerical values of the factors
$\displaystyle\frac{1}{n}
\sum_{j=1}^{n-2}\left[\vec{u}^{{}^{(\pm)}}_j
\cdot(\vec{D}^{F}\pm\vec{D}^{N})\right]^2$,
$\displaystyle\frac{1}{n}
\left[\vec{u}^{{}^{(\pm)}}_{n-1}
\cdot(\vec{D}^{F}\pm\vec{D}^{N})\right]^2$ and
$\displaystyle\frac{1}{n}
\left[\vec{u}^{{}^{(\pm)}}_{n}
\cdot(\vec{D}^{F}\pm\vec{D}^{N})\right]^2$
are all comparable, as can be seen from Table \ref{table}.
Furthermore, it turns out that the third term in
Eq.~(\ref{chi6}) is negligible compared to the second term.
This is because
\begin{eqnarray}
\Lambda^{{}^{(-)}}_-/n
\simeq0.46\sigma^2_{\text{dB}}\ll
\Lambda^{{}^{(-)}}_+/n
\simeq24\sigma^2_{\text{cal}}
\nonumber
\end{eqnarray}
are satisfied for our reference values (See Appendix~\ref{appendix4})
and because the values of
$\displaystyle\frac{1}{n}
\left[\vec{u}^{{}^{(-)}}_{n-1}
\cdot(\vec{D}^{F}-\vec{D}^{N})\right]^2$ and
$\displaystyle\frac{1}{n}
\left[\vec{u}^{{}^{(-)}}_{n}
\cdot(\vec{D}^{F}-\vec{D}^{N})\right]^2$
are comparable.
Thus we obtain
\begin{eqnarray}
\hspace*{-20mm}
\chi^2&\simeq&\frac{\sin^42\theta_{13}}{2}
\left\{
\frac{1}{\sigma^2_{\text{db}}/n}\,\cdot\,\frac{1}{n}
\sum_{j=1}^{n-2}
\left[\vec{u}^{{}^{(-)}}_j\cdot(\vec{D}^{F}-\vec{D}^{N})\right]^2\right.
\nonumber\\
&{\ }&\qquad\qquad+\left.\frac{1}{\sigma^2_{\text{db}}/n+0.46\sigma^2_{\text{dB}}}
\,\cdot\,\frac{1}{n}
\left[\vec{u}^{{}^{(-)}}_{n-1}\cdot(\vec{D}^{F}-\vec{D}^{N})\right]^2
\right\}.
\label{chi7}
\end{eqnarray}
This is the main result of this paper and is to be compared
with the result (\ref{chi4}) by the rate analysis.
 Once the baselines $L_F$ and $L_N$ are given,
then (\ref{chi7}) contains only $\sigma^2_{\text{db}}/\sqrt{n}$
and $\sigma^2_{\text{dB}}$ as the parameters
because the coefficients
$\displaystyle\frac{1}{n} \sum_{j=1}^{n-2}
\left[\vec{u}^{{}^{(-)}}_j\cdot(\vec{D}^{F}-\vec{D}^{N})\right]^2$
and 
$\displaystyle\frac{1}{n}
\left[\vec{u}^{{}^{(-)}}_{n-1}\cdot(\vec{D}^{F}-\vec{D}^{N})\right]^2$
are almost independent of $n$.
 These coefficients seem to give good measures for the potential of an experiment
in spectral analysis almost independently of $n$
and the sizes of errors;
 The larger values of these measures means the better setup in principle.

From numerical calculations we find that the dependence of
the first term in Eq.~(\ref{chi7}) on $L_F$ and $L_N$ is quite different
from that of the second term
and the values of
$\displaystyle\frac{1}{n}
\sum_{j=1}^{n-2}\left[\vec{u}^{{}^{(-)}}_j\cdot(\vec{D}^{F}-\vec{D}^{N})\right]^2$
and
$\displaystyle\frac{1}{n}
\left[\vec{u}^{{}^{(-)}}_{n-1}\cdot(\vec{D}^{F}-\vec{D}^{N})\right]^2$
are plotted as functions of $L_F$ and $L_N$ in Figs.~\ref{fig1}
and  \ref{fig2}.
$\displaystyle\frac{1}{n}
\sum_{j=1}^{n-2}\left[\vec{u}^{{}^{(-)}}_j\cdot(\vec{D}^{F}-\vec{D}^{N})\right]^2$
of the first term in Eq.~(\ref{chi7}) has the maximum value 0.37
at $L_F=$10.6km and $L_N$=8.4km, while
$\displaystyle\frac{1}{n}
\left[\vec{u}^{{}^{(-)}}_{n-1}\cdot(\vec{D}^{F}-\vec{D}^{N})\right]^2$
of the second is extremely small for these baselines.%
\footnote{
The values of the two optimized baselines for
$(1/n)\left[\vec{u}^{{}^{(-)}}_{n-1}\cdot(\vec{D}^{F}-\vec{D}^{N})\right]^2$
depend on the number $n$ of bins, and the present
values are obtained for $n=16$.
This is because the more number of bins we have, the more local maxima
and minima we could observe, and in order to see more local maxima and minima
we should have longer baselines.  Even if we take $n>16$, however,
this set of the baselines $L_F=$10.6km and $L_N$=8.4km
gives the local maximum and can be regarded approximately as the
optimized set.}
Theoretically, therefore,
the first term is optimized for $L_F=$10.6km and $L_N$=8.4km.
In this case the sensitivity to $\sin^22\theta_{13}$ in the limit of infinite statistics
is given by
\begin{eqnarray}
\left(\sin^22\theta_{13}\right)_{\text{limit}}^{\text{sys~only}}
&\simeq&\sqrt{\frac{\left.\chi^2\right|_{\text{90\%CL}}\times2}{0.37}}
\,\frac{\sigma_{\text{db}}}{\sqrt{n}}
=\sqrt{\frac{2.7\times2}{0.37}}\,\frac{\sigma_{\text{db}}}{\sqrt{n}}
=3.8\frac{\sigma_{\text{db}}}{\sqrt{n}}\label{sens1}\\
&\qquad&\mbox{\rm for}~L_F=10.6\mbox{\rm km~and~} L_N=8.4\mbox{\rm km}.
\nonumber
\end{eqnarray}
 If $\sigma_{\text{db}}/\sqrt{n}$ is smaller than $\sigma_{\text{u}}$,
the sensitivity (\ref{sens1}) of the spectral analysis is
better than (\ref{lbound1}) of the rate analysis.
It should be noted, however, that this sensitivity is attained
only if the statistical error in each bin becomes negligible compared to the
systematic errors which are already assumed to be smaller than
${\cal O}(0.1)\%$.  In order for this to happen at $L_F=$10km, we would need at
least 100 kton$\cdot$yr even at the Kashiwazaki-Kariwa nuclear power
plant, so the optimization of the first term in (\ref{chi7}),
namely the optimization of the spectral analysis,
is not realistic for the investigation of the oscillation of
$|\Delta m^2_{13}|={\cal O}(10^{-3})\eV^2$.

Next, we consider to optimize the the second term in Eq.~(\ref{chi7}),
i.e.\ the term which roughly corresponds to the rate analysis.
This is achieved at $L_F=$2.1km and $L_N$=0km.
For these values of the baselines,
Eq.~(\ref{chi7}) becomes numerically
\begin{eqnarray}
\hspace*{-40mm}
\chi^2&\simeq&\frac{\sin^42\theta_{13}}{2}
\left(
\frac{3.7\times10^{-3}}{\sigma^2_{\text{db}}/n}
+\frac{4.1\times10^{-1}}{\sigma^2_{\text{db}}/n+0.46
\sigma^2_{\text{dB}}}\right)
\label{chi8}\\
&\qquad&\mbox{\rm for}~L_F=2.1\mbox{\rm km~and~} L_N=0\mbox{\rm km}.
\nonumber
\end{eqnarray}
If $\sigma^2_{\text{db}}/n\gg(1/100)\sigma^2_{\text{dB}}$ then
the second term on the right-hand side
in Eq.~(\ref{chi8}) is dominant.
 In this case,
the sensitivity is not so much different from that in rate analysis
because the second term is understood as a rate-like one.
On the other hand, if
$\sigma^2_{\text{db}}/n\ll(1/100)\sigma^2_{\text{dB}}$ then
the first term in Eq.~(\ref{chi8}) dominates the
sensitivity, and it is given by
\begin{eqnarray}
\left(\sin^22\theta_{13}\right)_{\text{limit}}^{\text{sys~only}}
&\sim&\sqrt{\frac{2.7\times2}{3.7\times10^{-3}}}\,
\frac{\sigma_{\text{db}}}{\sqrt{n}}
=37\frac{\sigma_{\text{db}}}{\sqrt{n}}\label{sens3}\\
&\qquad&\mbox{\rm if}~
\frac{\sigma^2_{\text{db}}}{n}\ll\frac{1}{100}\sigma^2_{\text{dB}}
\simeq \frac{1}{100}\sigma_{\text{u}}^2 \simeq {\cal O}\left((0.01\%)^2\right).
\nonumber
\end{eqnarray}
 Since the coefficient $4.1\times 10^{-1}$ of the second term in (\ref{chi8})
is much larger than the coefficient $3.7\times 10^{-3}$ of the first term,
$\sigma_{\text{db}}/\sqrt{n}$ must be much smaller than
$\sigma_{\text{u}}$ in order to use (\ref{sens3}) and to improve
the sensitivity significantly beyond the one in the rate analysis .
If the setup is optimized for the rate analysis,
therefore, it is not easy for the spectral analysis
to give us much better sensitivity
than what the rate analysis does.
On the other hand, since the difference between the coefficients
of the first and second terms in (\ref{chi8}) can be smaller
for the setup which deviates from the optimal one
for the rate analysis,
the spectral analysis could compensate the loss
of the sensitivity due to the deviation.

To illustrate the usefulness of the formula (\ref{chi7}),
let us see if it reproduces the result
in~\cite{Huber:2003pm},
where the number $n$ of the energy bins was assumed to be 62,
the energy interval was 1.8MeV$\le E_\nu\le$8.0MeV,
$L_F$=1.7km, $L_N$=0.17km, and
the reference values are $|\Delta m^2_{31}|=3.0\times10^{-3}$eV$^2$,
$\sigma_{\text{dB}}$=0.6\%,
and $\sigma_{\text{db}}$=0.1\% (the most optimistic case
in~\cite{Huber:2003pm}).  We found from numerical
calculations that if we perform the spectrum analysis
with the energy interval 1.8MeV$\le E_\nu\le$8.0MeV,
the number $n$=62 of bins, the baselines
$L_F$=1.7km and $L_N$=0.17km, then Eq.~(\ref{chi7}) is modified
for $|\Delta m^2_{31}|=3.0\times10^{-3}$eV$^2$ as
\begin{eqnarray}
\hspace*{-40mm}
\chi^2&\simeq&\frac{\sin^42\theta_{13}}{2}
\left(
\frac{6.8\times10^{-2}}{\sigma^2_{\text{db}}/62}
+\frac{3.3\times10^{-1}}{\sigma^2_{\text{db}}/62+0.95\sigma^2_{\text{dB}}}\right)
\label{chi9}\\
&\qquad&\mbox{\rm for}~L_F=1.7\mbox{\rm km~and~} L_N=0.17\mbox{\rm km},
\nonumber
\end{eqnarray}
where we have used the fact $\vec{v}^2/n=1.2\times10^3$ and
$\vec{v}\cdot\vec{h}/n=-7.7$ for 1.8MeV$\le E_\nu\le$8.0MeV
and $n$=62 to derive the factor 0.95.
Note that the ratio of the coefficient 0.068 of the first term
to that 0.33 of the second term is bigger than that in Eq.~(\ref{chi8})
because $L_F$=1.7km is longer than the optimized baseline
$L\simeq$1.4km for the rate analysis in the case of
$|\Delta m^2_{31}|=3.0\times10^{-3}$eV$^2$.
This can be seen from Fig.~\ref{fig3}, where the value of
$\displaystyle\frac{1}{n}\sum_{j=1}^{n-2}
\left[\vec{u}^{{}^{(-)}}_j\cdot(\vec{D}^{F}-\vec{D}^{N})\right]^2$
increases for $L$ larger than the optimized baseline ($\simeq$2km)
for the rate analysis, i.e.\ the baseline which optimizes
$\displaystyle\frac{1}{n}
\left[\vec{u}^{{}^{(-)}}_{n-1}\cdot(\vec{D}^{F}-\vec{D}^{N})\right]^2$,
in the case of $|\Delta m^2_{31}|=2.2\times10^{-3}$eV$^2$.
Substituting the reference values $\sigma_{\text{dB}}$=0.6\%,
$\sigma_{\text{db}}$=0.1\% in Eq.~(\ref{chi9}) and the
value 2.7 of $\chi^2$ at the 90\%CL for one degree
of freedom, we have
\begin{eqnarray}
2.7&\simeq&\frac{\left(\sin^22\theta_{13}\right)_{\text{limit}}^{\text{sys~only}}}{2}
\left(\frac{6.8\times10^{-2}\times62}{10^{-6}}
+\frac{0.33}{0.95\times(6\times10^{-3})^2}\right)\nonumber\\
&\simeq&4.2\times10^6\times
\frac{\left(\sin^22\theta_{13}\right)_{\text{limit}}^{\text{sys~only}}}{2}
\nonumber
\end{eqnarray}
which gives
\begin{eqnarray}
\left(\sin^22\theta_{13}\right)_{\text{limit}}^{\text{sys~only}}
\simeq\sqrt{\frac{5.4}{4.2}}\times10^{-3}\sim 1\times10^{-3}.
\nonumber
\end{eqnarray}
This is almost consistent with the result in~\cite{Huber:2003pm}.
From this calculation it is obvious that
such a good sensitivity was obtained
in~\cite{Huber:2003pm} because the uncorrelated
bin-to-bin systematic error $\sigma_{\text{db}}$ was assumed
to be very small and the number of bins was large.

\section{Oscillation experiments for $\Delta m^2\sim {\cal O}(1)$eV$^2$}
In the previous section, in the case of reactor experiments
to measure $\sin^22\theta_{13}$ for $\Delta m^2\sim {\cal O}(10^{-3})$eV$^2$,
we have seen that the optimal baseline lengths for the spectral analysis
are very long and it is not a realistic idea to accumulate large number of events
at such baselines.
If we want to observe neutrino oscillations
for $\Delta m^2\sim {\cal O}(1)$eV$^2$, however, the optimal setup
for the spectral analysis becomes realistic.
With the standard three flavor scenario, there is little motivation
to look for neutrino oscillations for $\Delta m^2\sim {\cal O}(1)$eV$^2$,
but if we consider more general frameworks with sterile neutrinos
then oscillations with $\Delta m^2\sim {\cal O}(1)$eV$^2$ are still
possible.  In this section we briefly discuss such a possibility
for the sake of completeness.

In ref.~\cite{Sorel:2003hf} it was shown that the scheme with two
sterile neutrinos is still consistent with all the data,
and the best fit values for $\Delta m^2_{14}$ and
$\sin^22\theta_{14}$ are 0.9eV$^2$ and 0.058.
$\Delta m^2$=0.9eV$^2$ is the value for which the hypothesis
for neutrino oscillations is excluded only weakly by the Bugey
experiment~\cite{Declais:1994su}.  If the (3+2)--scheme is realized by Nature, therefore,
we should be able to observe neutrino oscillation at
$\Delta m^2\simeq$0.9eV$^2$.
By rescaling the baselines used in (\ref{sens1})
by $2.2\times10^{-3}$eV$^2$/0.9eV$^2\simeq
1/410$, we see in the limit of infinite statistics that
the optimum set of baselines for $\Delta m^2=0.9$eV$^2$ is
\begin{eqnarray}
L_F=26\text{m},~~L_N=21\text{m}.
\nonumber
\end{eqnarray}
If we put the two detectors with this set of baselines,
then in the limit of infinite statistics, assuming the
number of bins $n=25$, the bin-to-bin uncorrelated systematic error
$\sigma_{\text{db}}$=0.5\% and the mixing angle
$\sin^22\theta=0.058$, we could have
\begin{eqnarray}
\chi^2\simeq\frac{\sin^42\theta}{2}\cdot
\frac{0.37}{(5\times10^{-3})^2/25}\simeq 6\times10^2.
\label{sterile1}
\end{eqnarray}
Eq.~(\ref{sterile1}) would imply that neutrino oscillation for
$\Delta m^2\simeq$0.9eV$^2$ can be established
with tremendous significance.
This result is to be compared to the rate analysis with (\ref{chi4}),
where the set of optimum baselines for $\Delta m^2\simeq$0.9eV$^2$ would be
\begin{eqnarray}
L_F=5\text{m},~~L_N=0\text{m}.
\label{sterile2}
\end{eqnarray}
 Assuming that these baselines are technically possible,
$\chi^2$ at $\sin^22\theta=0.058$ would be
\begin{eqnarray}
\chi^2\simeq\frac{\sin^42\theta}{2}\cdot
\frac{0.82}{(0.6\times10^{-2})^2}\simeq 4\times10^1.
\label{sterile3}
\end{eqnarray}
From practical point of view, however, Eq.~(\ref{sterile2})
would imply that both the detectors be placed inside of the reactor
and Eq.~(\ref{sterile2}) is not realistic.
 The value of $\chi^2$ would be smaller than Eq.~(\ref{sterile3})
and much smaller than Eq.~(\ref{sterile1}).
 Thus, the spectral analysis is much more advantageous than
the rate analysis in search of neutrino oscillations
for $\Delta m^2\sim {\cal O}(1)$eV$^2$.

\section{Discussion and Conclusion}
 We applied our analytical method to estimate
the sensitivity to $\sin^22\theta$ in the limit of infinite statistics,
which we refer to as ``the systematic limit'',
in the spectral analysis at a reactor neutrino oscillation experiment with
one reactor and two detectors.
In a rate analysis, the systematic limit (\ref{sens0}) is 
set by the the uncorrelated systematic error $\sigma_{\text{u}}$
of the detectors, and is given by 2.8$\sigma_{\text{u}}$~\cite{Sugiyama:2004bv}.
In a spectrum analysis, we derived an analogous formula for
$\chi^2$ which gives the systematic limit.
The formula for $\chi^2$ contains
two terms which could potentially dominate
$\chi^2$ (\ref{chi7}).  One is the contribution
from the bin-to-bin correlated detector specific
systematic error $\sigma_{\text{dB}}$
and the other is the one from $\sigma_{\text{db}}/\sqrt{n}$
which is the uncorrelated bin-to-bin systematic error
divided by the square-root of the number $n$ of bins.
 The optimal setup for the spectral analysis is found
to be unrealistic for $\Delta m^2 = {\cal O}(10^{-3})\eV^2$
because the optimal baselines are too long to accumulate
large number of events.
 The optimization can be realistic, however,
for $\Delta m^2 = {\cal O}(1)\eV^2$
to search for sterile neutrino oscillations.
 If the baselines are optimized for the rate analysis,
the improvement of the sensitivity with the spectral analysis
is not significant in most cases.
On the other hand,
the spectral analysis can be effective for the experiment
whose setup deviates from the optimal one for the rate analysis.
 For example,
the spectral analysis could compensate the disadvantage
of the rate analysis in the KASKA experiment which
has near detectors with relatively long baselines ($\sim$ 300--400m),
or that in the DCHOOZ experiment that has a far detector
at a distance ($\simeq 1$km) shorter than the optimal one
for the rate analysis.
 In any case,
it is very important in the
future reactor experiments to estimate $\sigma_{\text{dB}}$
and especially $\sigma_{\text{db}}$.

\appendix

\section{Derivation of the covariance matrix\label{appendix1}}
After taking the limit $1/t^A_i\rightarrow0$
of infinite statistics,
Eq.~(\ref{chipull}) becomes
\begin{eqnarray}
\hspace*{-20mm}
\displaystyle
\chi^2&=&\min_{\alpha's}\Bigg\{
\displaystyle\sum_{A=N,F}\sum_{i=1}^n
\frac{ \left(m^A_i/t^A_i-1-\alpha-\alpha^A-\alpha_i-\alpha_{\text{cal}}^A v^A_i
\right)^2 }{ (\sigma^A_{\text{db}})^2 }\nonumber\\
&+&\displaystyle\sum_{A=N,F}\left[\left(
\frac{\alpha^A} {\sigma_{\text{dB}}}\right)^2
+\left(\frac{\alpha_{\text{cal}}^A} {\sigma_{\text{cal}}}\right)^2\right]
+\displaystyle\sum_{i=1}^n\left(
\frac{\alpha_i} {\sigma_{\text{Db}}}\right)^2
+\left(\frac{\alpha} {\sigma_{\text{DB}}}\right)^2\Bigg\}\nonumber\\
&\equiv& \min_{\alpha's} \chi^2_\alpha. \nonumber
\end{eqnarray}
Let us redefine the variables
\begin{eqnarray}
\hspace*{-20mm}
y^A_i&\equiv&\frac{m^A_i-t^A_i} {t^A_i}
= -\sin^22\theta_{13} D^A_i
\nonumber
\end{eqnarray}
and let us introduce a vector notation
\begin{eqnarray}
\hspace*{-20mm}
\vec{y}^{N}\equiv\left(\begin{array}{c}
y^N_1\\
\vdots\\
y^N_n
\end{array}\right),\qquad
\vec{y}^{F}\equiv\left(\begin{array}{c}
y^F_1\\
\vdots\\
y^F_n
\end{array}\right).
\nonumber
\end{eqnarray}
Following the discussions in the Appendix A in~\cite{Sugiyama:2004bv},
the matrix element of the covariance matrix
can be obtained as the expectation value of $y^A_i\,y^B_j$:
\begin{eqnarray}
\hspace*{-20mm}
\left(\rho\right)^{AB}_{ij}&=&
\left\langle y^A_i y^B_j \right\rangle\nonumber\\
&\equiv&{\cal N}
\int d\alpha\prod_{A=N,F}\int d\vec{y}^A\int d\alpha^A\int d\alpha_{\text{cal}}^A
\prod_{i=1}^n\int d\alpha_i~y^A_i\,y^B_j\,
\exp\left(-\frac{\chi^2_\alpha}{2}\right),
\nonumber
\end{eqnarray}
where the normalization ${\cal N}$ is defined in such a way
that $\langle 1 \rangle=1$.  From a straightforward calculation,
we have
\begin{eqnarray}
\hspace*{-20mm}
\left\langle y^A_i y^B_j \right\rangle
&=&\delta^{AB}\delta_{ij}\,\sigma_{\text{db}}^2
+\sigma_{\text{DB}}^2
+\delta^{AB}\,\sigma_{\text{dB}}^2
+\delta_{ij}\,\sigma_{\text{Db}}^2
+\delta^{AB}v^A_i v^A_j \,\sigma^2_{\text{cal}}.
\nonumber
\end{eqnarray}
Thus we have
\begin{eqnarray}
\hspace*{-20mm}
\displaystyle
\chi^2=
\left(\begin{array}{ll}
\vec{y}^{N\ T},&\vec{y}^{F\ T}
\end{array}
\right)
\rho^{-1}
\left(\begin{array}{l}
\vec{y}^{N}\\
\vec{y}^{F}
\end{array}\right)
=\sin^42\theta_{13} \left(\begin{array}{ll}
\vec{D}^{N\ T},&\vec{D}^{F\ T}
\end{array}
\right)
\rho^{-1}
\left(\begin{array}{l}
\vec{D}^{N}\\
\vec{D}^{F}
\end{array}\right),
\nonumber
\end{eqnarray}
where the covariance matrix $\rho$ is defined by
\begin{eqnarray}
\hspace*{-20mm}
\rho=\left(\begin{array}{ll}
M&N\\
N&M
\end{array}\right),
\label{rho}
\end{eqnarray}
with
\begin{eqnarray}
\hspace*{-20mm}
M&\equiv& \left(\sigma^2_{\text{Db}}+\sigma^2_{\text{db}}\right)I_n
+\left(\sigma^2_{\text{DB}}+\sigma^2_{\text{dB}}\right)H_n
+\sigma^2_{\text{cal}}G_n,\nonumber\\
N&\equiv& \sigma^2_{\text{Db}}I_n
+\sigma^2_{\text{DB}}H_n.
\label{n}
\end{eqnarray}
Here $I_n$ is an $n\times n$ unit matrix,
$H_n$ and $G_n$ are $n\times n$ matrices defined by
\begin{eqnarray}
\hspace*{-20mm}
H_n&\equiv&\left(\begin{array}{ccc}
1&\cdots&1\\
\vdots&&\vdots\\
1&\cdots&1
\end{array}\right),\label{hn}\\
G_n&\equiv&\left(\begin{array}{cccc}
v_1^2&v_1v_2&\cdots&v_1v_n\\
v_1v_2&v_2^2&\cdots&v_2v_n\\
\vdots&\vdots&&\vdots\\
v_1v_n&v_2v_n&\cdots&v_n^2
\end{array}\right),
\label{gn}
\end{eqnarray}
where $v_j~(j=1,\cdots,n)$ is defined by Eqs.~(\ref{v}),
and (\ref{v3}).
 Note that the covariance matrix does not include
oscillation parameters but only errors.
 Diagonalization of the covariance matrix is useful
to see which errors dominate $\chi^2$.

Now (\ref{rho}) can be cast into a block diagonal by
\begin{eqnarray}
\hspace*{-20mm}
\left(\begin{array}{ll}
M&N\\
N&M
\end{array}\right)
=
\frac{1} {2}
\left(\begin{array}{rr}
I_n&-I_n\\
I_n&I_n
\end{array}\right)
\left(\begin{array}{cc}
M+N&0\\
0&M-N
\end{array}\right)
\left(\begin{array}{rr}
I_n&I_n\\
-I_n&I_n
\end{array}\right).
\nonumber
\end{eqnarray}
We obtain
\begin{eqnarray}
\hspace*{-5mm}
\chi^2=\frac{\sin^42\theta_{13}}{2}
\left(\begin{array}{ll}
\vec{D}^{F\ T}+\vec{D}^{N\ T},&\vec{D}^{F\ T}-\vec{D}^{N\ T}
\end{array}
\right)
\left(\begin{array}{cc}
M+N&0\\
0&M-N
\end{array}\right)^{-1}
\left(\begin{array}{l}
\vec{D}^{F}+\vec{D}^{N}\\
\vec{D}^{F}-\vec{D}^{N}
\end{array}\right).
\label{chi5}
\end{eqnarray}

We can formally diagonalize the symmetric matrices $M+N$ and $M-N$
assuming the existence of $(n-2)$ orthonormal eigenvectors which are
orthogonal to $\vec{v}$ and $\vec{h}\equiv\left(1,\cdots,1\right)^T$.
This is described in Appendix~\ref{appendix2}.
Now $M+N$ and $M-N$ are diagonalized by the
orthogonal matrices $U^{{}^{(+)}}$ and $U^{{}^{(-)}}$:
\begin{eqnarray}
\hspace*{-20mm}
M+N&=&U^{{}^{(+)}}{\cal D}^{{}^{(+)}}U^{{}^{(+)}\ T},\nonumber\\
M-N&=&U^{{}^{(-)}}{\cal D}^{{}^{(-)}}U^{{}^{(-)}\ T},
\label{diagmn}
\end{eqnarray}
where
\begin{eqnarray}
\hspace*{-20mm}
{\cal D}^{{}^{(\pm)}}\equiv\mbox{\rm diag}\left(
\lambda^{{}^{(\pm)}}_1,\cdots,\lambda^{{}^{(\pm)}}_n\right)
\nonumber
\end{eqnarray}
and $\lambda^{{}^{(\pm)}}_j$ are the eigenvalues of $M\pm N$.
From (\ref{chi5}) and (\ref{diagmn}) we obtain
\begin{eqnarray}
\chi^2&=&\frac{\sin^42\theta_{13}}{2}
\left(\begin{array}{ll}
(\vec{D}^{F\ T}+\vec{D}^{N\ T})U^{{}^{(+)}},&
(\vec{D}^{F\ T}-\vec{D}^{N\ T})U^{{}^{(-)}}
\end{array}
\right)
\nonumber\\&{\ }&\times
\left(\begin{array}{cc}
{\cal D}^{{}^{(+)}}&0\\
0&{\cal D}^{{}^{(-)}}
\end{array}\right)^{-1}
\left(\begin{array}{l}
U^{{}^{(+)}\ T}(\vec{D}^{F}+\vec{D}^{N})\\
U^{{}^{(-)}\ T}(\vec{D}^{F}-\vec{D}^{N})
\end{array}\right)\nonumber\\
&=&\frac{\sin^42\theta_{13}}{2}
\left\{\sum_{j=1}^{n}
\frac{\left[\vec{u}^{{}^{(+)}}_j\cdot(\vec{D}^{F}+\vec{D}^{N})
\right]^2}{\lambda^{{}^{(+)}}_j}
+\sum_{j=1}^{n}
\frac{\left[\vec{u}^{{}^{(-)}}_j\cdot(\vec{D}^{F}-\vec{D}^{N})
\right]^2}{\lambda^{{}^{(-)}}_j}
\right\},
\nonumber
\end{eqnarray}
where $\vec{u}^{{}^{(\pm)}}_j$ are the orthonormal eigenvectors
which appear in $U^{{}^{(\pm)}}$:
\begin{eqnarray}
\hspace*{-20mm}
U^{{}^{(\pm)}}\equiv\left(\vec{u}^{{}^{(\pm)}}_1,\cdots,
\vec{u}^{{}^{(\pm)}}_n\right).
\nonumber
\end{eqnarray}
Putting
\begin{eqnarray}
\hspace*{-20mm}
M\pm N&\equiv&a^{{}^{(\pm)}}I_n+b^{{}^{(\pm)}}H_n+cG_n,
\nonumber
\end{eqnarray}
we have from (\ref{n})
\begin{eqnarray}
\hspace*{-20mm}
a^{{}^{(+)}}&=&\sigma^2_{\text{db}}+2\sigma^2_{\text{Db}},\nonumber\\
b^{{}^{(+)}}&=&\sigma^2_{\text{dB}}+2\sigma^2_{\text{DB}},\nonumber\\
a^{{}^{(-)}}&=&\sigma^2_{\text{db}},\nonumber\\
b^{{}^{(-)}}&=&\sigma^2_{\text{dB}},\nonumber\\
c&=&\sigma^2_{\text{cal}}.
\nonumber
\end{eqnarray}
We will see in Appendix~\ref{appendix2} that $\lambda^{{}^{(\pm)}}_j$
are given by
\begin{eqnarray}
\hspace*{-20mm}
\lambda^{{}^{(+)}}_j&=&\left\{
\begin{array}{l}
a^{{}^{(+)}}\qquad\qquad\quad(j=1,\cdots,n-2)\\
a^{{}^{(+)}}+\Lambda^{{}^{(+)}}_-\qquad(j=n-1)\\
a^{{}^{(+)}}+\Lambda^{{}^{(+)}}_+\qquad(j=n)
\end{array},\right.
\nonumber\\
\lambda^{{}^{(-)}}_j&=&\left\{
\begin{array}{l}
a^{{}^{(-)}}\qquad\qquad\quad(j=1,\cdots,n-2)\\
a^{{}^{(-)}}+\Lambda^{{}^{(-)}}_-\qquad(j=n-1)\\
a^{{}^{(-)}}+\Lambda^{{}^{(-)}}_+\qquad(j=n)
\end{array},\right.
\nonumber
\end{eqnarray}
where $\Lambda^{{}^{(\pm)}}_\pm$ are given by
\begin{eqnarray}
\hspace*{-20mm}
\Lambda^{{}^{(+)}}_\pm\equiv
\frac{b^{{}^{(+)}}n+c\,\vec{v}^2}{2}\pm
\sqrt{\left(\frac{b^{{}^{(+)}}n-c\,\vec{v}^2}{2}\right)^2
+b^{{}^{(+)}}c\left(\vec{v}\cdot\vec{h}\right)^2},
\nonumber\\
\Lambda^{{}^{(-)}}_\pm\equiv
\frac{b^{{}^{(-)}}n+c\,\vec{v}^2}{2}\pm
\sqrt{\left(\frac{b^{{}^{(-)}}n-c\,\vec{v}^2}{2}\right)^2
+b^{{}^{(-)}}c\left(\vec{v}\cdot\vec{h}\right)^2}.
\nonumber
\end{eqnarray}
Hence we obtain the $\chi^2$ (\ref{chi6}).
In Appendix~\ref{appendix4} we will see that $\Lambda^{{}^{(\pm)}}_\pm/n$ are
approximately independent of the number $n$ of bins for
the energy interval 2.8MeV$\le E_\nu\le$7.8MeV.

\section{Diagonalization of the $n\times n$ matrix\label{appendix2}}
In this appendix we show how we diagonalize the
matrix
\begin{eqnarray}
aI_n+bH_n+cG_n,
\nonumber
\end{eqnarray}
where $I_n$ is the $n\times n$ unit matrix,
and $H_n$, $G_n$ and $v_j$ are given by Eqs.~(\ref{hn}), (\ref{gn})
and Eq.~(\ref{v}), respectively.
Since $aI_n$ only shifts each eigenvalue by $a$,
we will discuss the diagonalization of $bH_n+cG_n$ here.
Let us introduce the notation
\begin{eqnarray}
\vec{h}\equiv
\left(\begin{array}{c}
1\\
\vdots\\
1
\end{array}\right).
\nonumber
\end{eqnarray}
When $\vec{h}\cdot\vec{v}\ne0$, which is satisfied
in the present case for Eq.~(\ref{v}), we can always
find in $n$-dimensional Euclidean space
$n-2$ orthonormal eigenvectors $\vec{u}_j~(j=1,\cdots,n-2)$
which are orthogonal to $\vec{h}$ and $\vec{v}$, i.e.,
\begin{eqnarray}
\vec{h}\cdot\vec{u}_j=\vec{v}\cdot\vec{u}_j=0\qquad(j=1,\cdots,n-2).
\label{b1}
\end{eqnarray}
From Eq.~(\ref{b1}) we have
\begin{eqnarray}
H_n\vec{u}_j=\vec{u}_j\cdot\vec{h}
\left(\begin{array}{c}
1\\
\vdots\\
1
\end{array}\right)=0,
\nonumber\\
G_n\vec{u}_j=\vec{u}_j\cdot\vec{v}
\left(\begin{array}{c}
v_1\\
\vdots\\
v_n
\end{array}\right)=0.
\label{b2}
\end{eqnarray}
Eq.~(\ref{b2}) indicates that $\vec{u}_j~(j=1,\cdots,n-2)$ are
the $n-2$ eigenvectors of $bH_n+cG_n$ with the eigenvalue 0:
\begin{eqnarray}
\left(bH_n+cG_n\right)\vec{u}_j=0\qquad(j=1,\cdots,n-2).
\nonumber
\end{eqnarray}
It turns out that
the remaining 2 eigenvectors can be constructed
out of $\vec{h}$ and $\vec{v}$.
Straightforward calculations show the following:
\begin{eqnarray}
H_n\vec{h}=n\vec{h},\nonumber\\
G_n\vec{h}=(\vec{v}\cdot\vec{h})\vec{v},\nonumber\\
H_n\vec{v}=(\vec{v}\cdot\vec{h})\vec{h},\nonumber\\
G_n\vec{v}=(\vec{v}^2)\vec{v}.
\nonumber
\end{eqnarray}
Hence we obtain
\begin{eqnarray}
\left(bH_n+cG_n\right)
\left(\begin{array}{c}
\vec{h}\\
\vec{v}
\end{array}\right)=
\left(\begin{array}{cc}
bn&c\,\vec{v}\cdot\vec{h}\\
b\vec{v}\cdot\vec{h}&c\,\vec{v}^2
\end{array}\right)
\left(\begin{array}{c}
\vec{h}\\
\vec{v}
\end{array}\right).
\nonumber
\end{eqnarray}
This suggests that multiplication of
the vectors $\sqrt{b}\,\vec{h}$ and
$\sqrt{c}\,\vec{v}$ by $bH_n+cG_n$
give the symmetric matrix
\begin{eqnarray}
\left(bH_n+cG_n\right)
\left(\begin{array}{c}
\sqrt{b}\,\vec{h}\\
\sqrt{c}\,\vec{v}
\end{array}\right)=
\left(\begin{array}{cc}
bn&\sqrt{bc}\,\vec{v}\cdot\vec{h}\\
\sqrt{bc}\,\vec{v}\cdot\vec{h}&c\,\vec{v}^2
\end{array}\right)
\left(\begin{array}{c}
\sqrt{b}\,\vec{h}\\
\sqrt{c}\,\vec{v}
\end{array}\right).
\label{b3}
\end{eqnarray}
The symmetric matrix on the right-hand side can be diagonalized as
\begin{eqnarray}
\left(\begin{array}{cc}
bn&\sqrt{bc}\,\vec{v}\cdot\vec{h}\\
\sqrt{bc}\,\vec{v}\cdot\vec{h}&c\,\vec{v}^2
\end{array}\right)
=\left(\begin{array}{cc}
\cos\psi&-\sin\psi\\
\sin\psi&\cos\psi
\end{array}\right)
\left(\begin{array}{cc}
\Lambda_+&0\\
0&\Lambda_-
\end{array}\right)
\left(\begin{array}{cc}
\cos\psi&\sin\psi\\
-\sin\psi&\cos\psi
\end{array}\right),
\nonumber
\end{eqnarray}
where $\psi$ is given by
\begin{eqnarray}
\tan2\psi\equiv\frac{\sqrt{bc}\,\vec{v}\cdot\vec{h}}
{\left(bn-c\,\vec{v}^2\right)/2}
\label{b45}
\end{eqnarray}
and the eigenvalues $\Lambda_{\pm}$ are
\begin{eqnarray}
\Lambda_\pm\equiv
\frac{bn+c\,\vec{v}^2}{2}\pm
\sqrt{\left(\frac{bn-c\,\vec{v}^2}{2}\right)^2
+\left(\sqrt{bc}\,\vec{v}\cdot\vec{h}\right)^2}.
\label{b4}
\end{eqnarray}
Therefore, if we define
\begin{eqnarray}
\left(\begin{array}{c}
\vec{w}_1\\
\vec{w}_2
\end{array}\right)
\equiv
\left(\begin{array}{cc}
\cos\psi&\sin\psi\\
-\sin\psi&\cos\psi
\end{array}\right)
\left(\begin{array}{c}
\sqrt{b}\,\vec{h}\\
\sqrt{c}\,\vec{v}
\end{array}\right),
\nonumber
\end{eqnarray}
then from Eq.~(\ref{b3}) we have
\begin{eqnarray}
\left(bH_n+cG_n\right)
\left(\begin{array}{c}
\vec{w}_1\\
\vec{w}_2
\end{array}\right)=
\left(\begin{array}{cc}
\Lambda_+&0\\
0&\Lambda_-
\end{array}\right)
\left(\begin{array}{c}
\vec{w}_1\\
\vec{w}_2
\end{array}\right).
\nonumber
\end{eqnarray}
It is easy to show that the normalizations of
$\vec{w}_1$ and $\vec{w}_2$ are $\sqrt{\Lambda}_+$
and $\sqrt{\Lambda}_-$, respectively.
 We see that the $(n-1)$-th and the $n$-th
orthonormal eigenvectors are given by
\begin{eqnarray}
\left(\begin{array}{c}
\vec{u}_{n-1}\\
\vec{u}_n
\end{array}\right)\equiv
\left(\begin{array}{c}
\vec{w}_2/\sqrt{\Lambda_-}\\
\vec{w}_1/\sqrt{\Lambda_+}
\end{array}\right)=
\left(\begin{array}{c}
\displaystyle\frac{1}{\sqrt{\Lambda_-}}\left(
-\sqrt{b}\,\vec{h}\sin\psi+\sqrt{c}\,\vec{v}\cos\psi\right)\\
\displaystyle\frac{1}{\sqrt{\Lambda_+}}\left(
\sqrt{b}\,\vec{h}\cos\psi+\sqrt{c}\,\vec{v}\sin\psi\right)
\end{array}\right)
\label{b5}
\end{eqnarray}
with the eigenvalues $\Lambda_-$ and $\Lambda_+$.

We note in passing that (\ref{b5}) has the correct behaviors
in the limit $b\ll c$ or $b\gg c$.
In the limit $b\ll c$ we have
\begin{eqnarray}
\left(\begin{array}{c}
\vec{u}_{n-1}\\
\vec{u}_n
\end{array}\right)\rightarrow
\left(\begin{array}{c}
\displaystyle-\frac{\vec{h}-(\vec{v}\cdot\vec{h})\vec{v}/\vec{v}^2}
{\sqrt{n-(\vec{v}\cdot\vec{h})^2/\vec{v}^2}}\\
\displaystyle\frac{\vec{v}}{\sqrt{\vec{v}^2}}
\end{array}\right)
\nonumber
\end{eqnarray}
with
\begin{eqnarray}
\left(\begin{array}{c}
\Lambda_-\\
\Lambda_+
\end{array}\right)\rightarrow
\left(\begin{array}{c}
\displaystyle b\left[n-
\frac{(\vec{v}\cdot\vec{h})^2}{\vec{v}^2}\right]\\
\displaystyle c\,\vec{v}^2
\end{array}\right),
\label{bllc}
\end{eqnarray}
while in the limit $b\gg c$ we obtain
\begin{eqnarray}
\left(\begin{array}{c}
\vec{u}_{n-1}\\
\vec{u}_n
\end{array}\right)\rightarrow
\left(\begin{array}{c}
\displaystyle\frac{\vec{v}-(\vec{v}\cdot\vec{h})\vec{h}/n}
{\sqrt{\vec{v}^2-(\vec{v}\cdot\vec{h})^2/n}}\\
\displaystyle\frac{\vec{h}}{\sqrt{n}}
\end{array}\right)
\nonumber
\end{eqnarray}
with
\begin{eqnarray}
\left(\begin{array}{c}
\Lambda_-\\
\Lambda_+
\end{array}\right)\rightarrow
\left(\begin{array}{c}
\displaystyle c\left[\vec{v}^2-
\frac{(\vec{v}\cdot\vec{h})^2}{n}\right]\\
\displaystyle bn
\end{array}\right).
\nonumber
\end{eqnarray}

Putting everything together, we obtain
\begin{eqnarray}
U^T\left(bH_n+cG_n\right)U=
\mbox{\rm diag}(0,\cdots,0,\Lambda_-,\Lambda_+)
\label{b6}
\end{eqnarray}
with
\begin{eqnarray}
U=\left(\vec{u}_1,\cdots,\vec{u}_n\right),
\nonumber
\end{eqnarray}
where $\vec{u}_1, \cdots, \vec{u}_{n-2}$ are $n-2$
orthonormal vectors which are orthogonal to
$\vec{v}$ and $\vec{h}$, $\vec{u}_{n-1}$ and
$\vec{u}_n$ are given by Eq.~(\ref{b5}), $\Lambda_-$ and $\Lambda_+$
are given by Eq.~(\ref{b4}).  Adding the unit matrix $aI_n$ to
Eq.~(\ref{b6}), we finally obtain
\begin{eqnarray}
U^T\left(aI_n+bH_n+cG_n\right)U=
\mbox{\rm diag}(a,\cdots,a,a+\Lambda_-,a+\Lambda_+).
\nonumber
\end{eqnarray}

\section{The properties of $\vec{v}$\label{appendix4}}
\begin{figure}
\vspace*{20mm}
\hspace*{-15mm}
\includegraphics[scale=1.4]{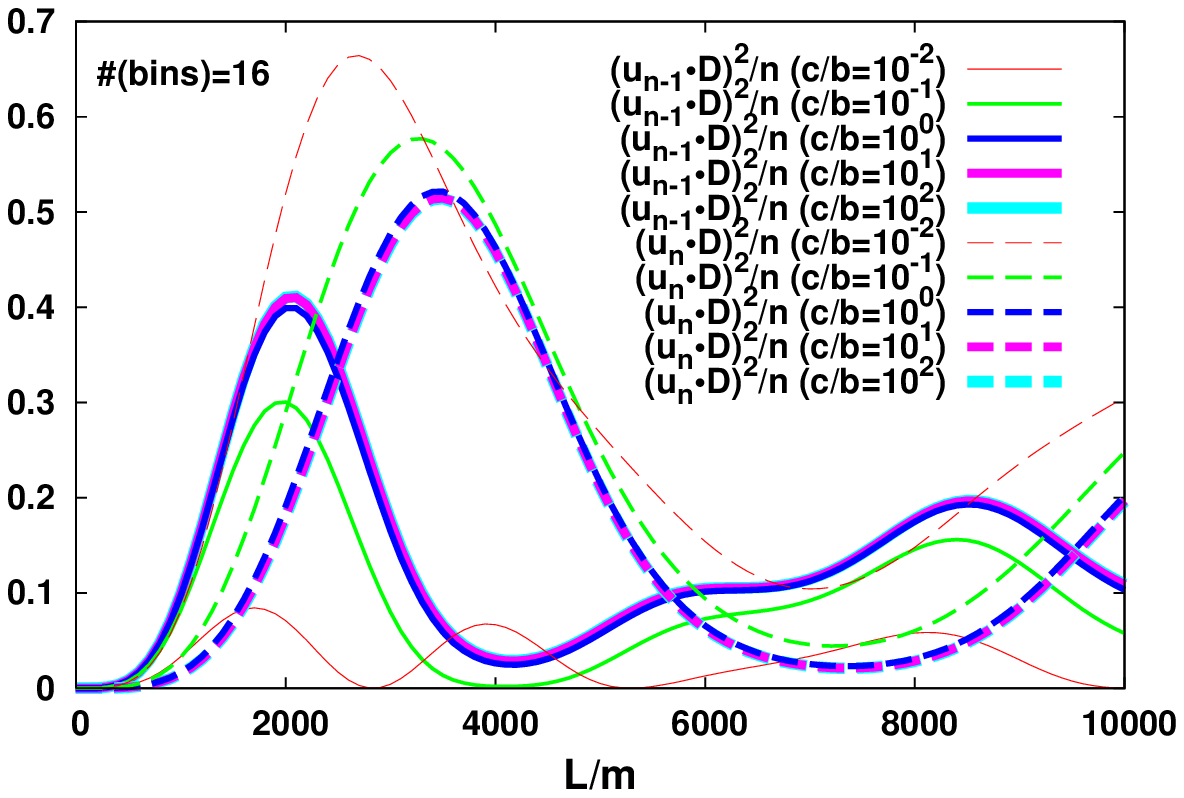}
\caption{
The values of
$\displaystyle\frac{1}{n}
\left(\vec{u}^{{}^{(-)}}_{n-1}\cdot\vec{D}^{F}\right)^2$
(the solid lines) and
$\displaystyle\frac{1}{n}
\left(\vec{u}^{{}^{(-)}}_n\cdot\vec{D}^{F}\right)^2$
(the dashed lines) for different values of
$c/b^{{}^{(-)}}$ at $L_F=L$, $L_N=0$,
where $b^{{}^{(-)}}=\sigma^2_{\text{dB}}=(0.6\%)^2$ and
$c=\sigma^2_{\text{cal}}$.
The reference values in the main text give
$c/b^{{}^{(-)}}\gtrsim 1$ and the two quantities
in the figure for $c/b^{{}^{(-)}}\gtrsim 1$
are approximated by the limit $c/b^{{}^{(-)}}\gg1$.
The number $n$ of bins is 16 and
$|\Delta m^2_{13}|=2.2\times10^{-3}$eV$^2$.
}
\label{fig4}
\end{figure}

\begin{figure}
\vspace*{-20mm}
\hspace*{-30mm}
\includegraphics[scale=1.6]{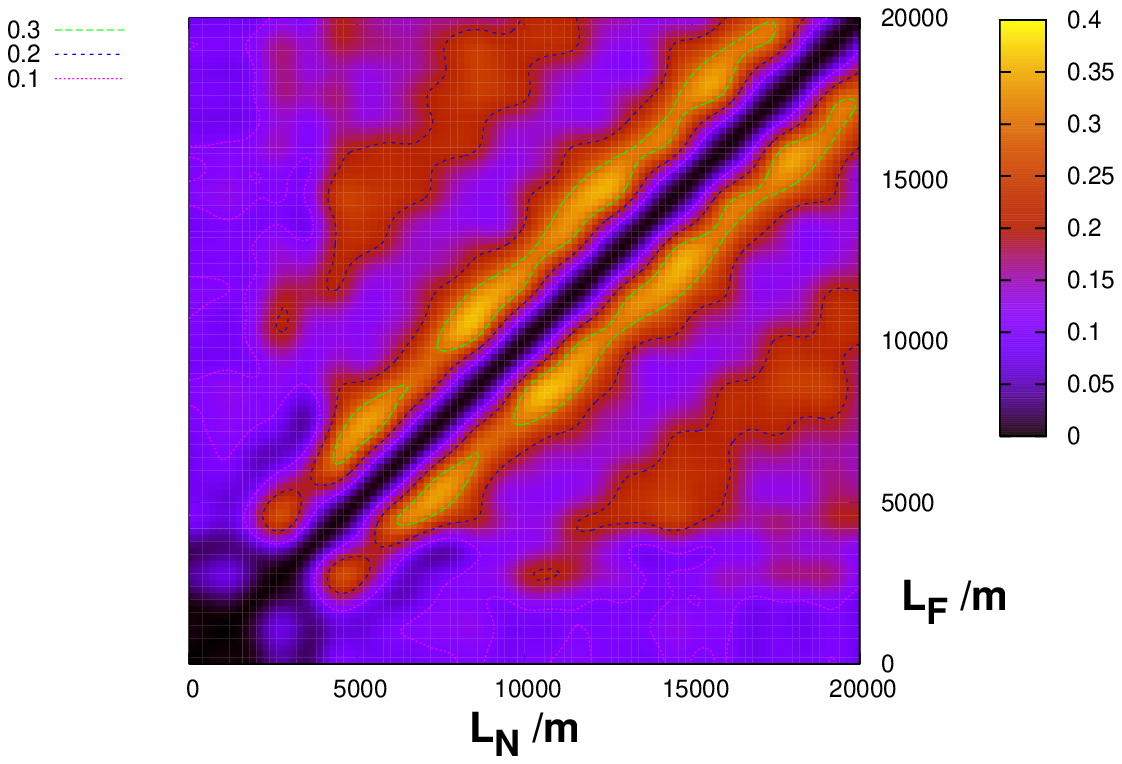}
\caption{
The values of
$\displaystyle\frac{1}{n}
\sum_{j=1}^{n-2}\left[\vec{u}^{{}^{(-)}}_j
\cdot(\vec{D}^{F}-\vec{D}^{N})\right]^2$
as a function of $L_F$ and $L_N$.
It has the maximum value 0.37 at
$L_F$=10.6km and $L_N$=8.4km.
The number $n$ of bins is 16 and
$|\Delta m^2_{13}|=2.2\times10^{-3}$eV$^2$.
}
\label{fig1}
\end{figure}

\begin{figure}
\vspace*{-20mm}
\hspace*{-30mm}
\includegraphics[scale=1.6]{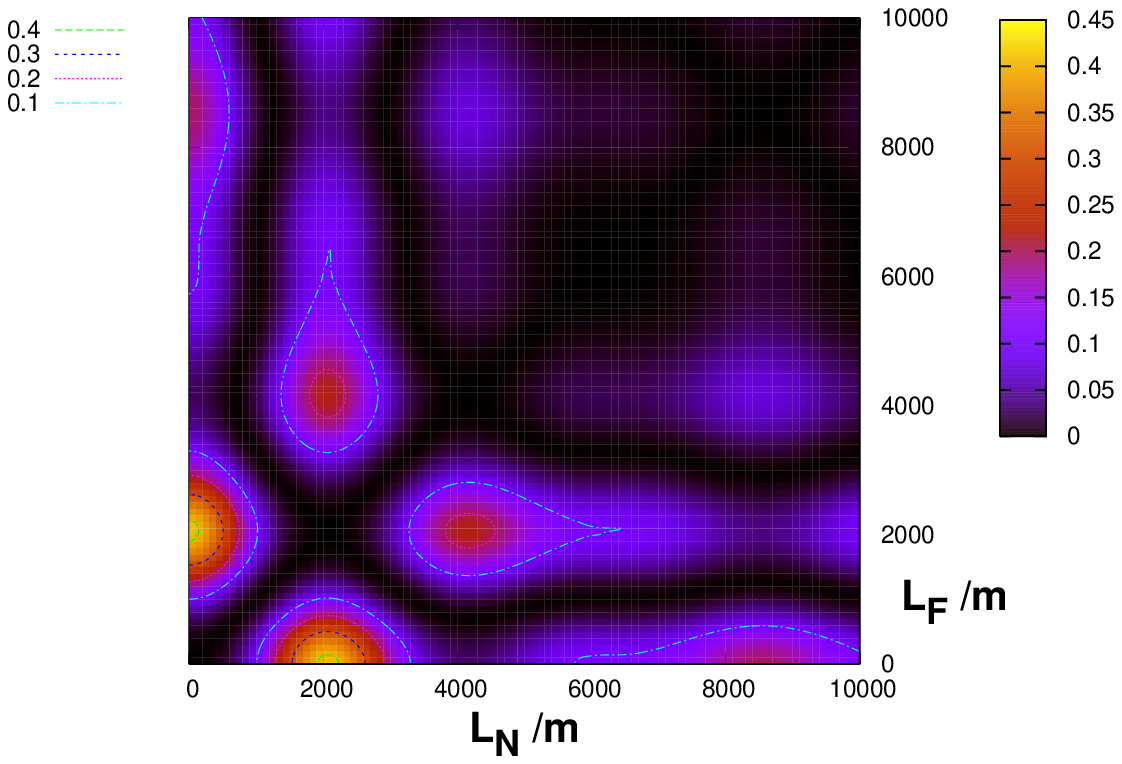}
\caption{
The values of
$\displaystyle\frac{1}{n}
\left[\vec{u}^{{}^{(-)}}_{n-1}\cdot(\vec{D}^{F}-\vec{D}^{N})\right]^2$
as a function of $L_F$ and $L_N$.
It has the maximum value 0.41 at
$L_F$=2.1km and $L_N$=0km.
The number $n$ of bins is 16 and
$|\Delta m^2_{13}|=2.2\times10^{-3}$eV$^2$.
}
\label{fig2}
\end{figure}

From numerical calculations, we find
for $n\gtrsim 8$ and for 2.8MeV$\le E_\nu\le$7.8MeV that
\begin{eqnarray}
\frac{1}{n}\vec{v}\cdot\vec{h}&\simeq&-3.6=
\mbox{\rm independent of}~n,\nonumber\\
\frac{1}{n}\vec{v}^2&\simeq&24=
\mbox{\rm independent of}~n.\nonumber
\end{eqnarray}
If we introduce the rescaled variables
\begin{eqnarray}
\vec{\tilde{v}}&\equiv&\frac{1}{\sqrt{n}}\vec{v},\nonumber\\
\vec{\tilde{h}}&\equiv&\frac{1}{\sqrt{n}}\vec{h},
\nonumber
\end{eqnarray}
then $\vec{\tilde{v}}^2$ and $\vec{\tilde{v}}\cdot\vec{\tilde{h}}$
are almost independent of the number $n$ of bins.
Furthermore, we rescale other quantities as follows:
\begin{eqnarray}
\tilde{\Lambda}_\pm&\equiv&n\lambda_\pm,
\nonumber
\end{eqnarray}
where $\Lambda_\pm$ is given by Eq.~(\ref{b4}).
$\tilde{\Lambda}_\pm$ is approximately independent of $n$
because $\vec{\tilde{v}}^2$ and $\vec{\tilde{v}}\cdot\vec{\tilde{h}}$
are:
\begin{eqnarray}
\tilde{\Lambda}_\pm&=&
\frac{b+c\,\vec{\tilde{v}}^2}{2}\pm
\sqrt{\left(\frac{b-c\,\vec{\tilde{v}}^2}{2}\right)^2
+bc\left(\vec{\tilde{v}}\cdot\vec{\tilde{h}}\right)^2}\nonumber\\
&\simeq&
\frac{b+24c}{2}\pm
\sqrt{\left(\frac{b-24c}{2}\right)^2
+3.6^2bc}.
\label{tlambda}
\end{eqnarray}
Eq.~(\ref{tlambda}) suggests that
we can expand $\tilde{\Lambda}_\pm$ in either $b$ or $c$
depending on whether $b\ll c$ or $b\gg c$.
In the present case, our reference values for the $(-)$ sector
are
\begin{eqnarray}
b&\equiv& b^{{}^{(-)}}=\sigma^2_{\text{dB}}\simeq{\cal O}((0.1\%)^2),
\nonumber\\
c&=&\sigma^2_{\text{cal}}={\cal O}((0.1\%)^2).
\nonumber
\end{eqnarray}
 Fig.~\ref{fig4} shows the values of
$\displaystyle\frac{1}{n}\sum_{j=1}^{n-2}\left[\vec{u}^{{}^{(\pm)}}_j
\cdot(\vec{D}^{F}\pm\vec{D}^{N})\right]^2$,
$\displaystyle\frac{1}{n}\left[\vec{u}^{{}^{(\pm)}}_{n-1}
\cdot(\vec{D}^{F}\pm\vec{D}^{N})\right]^2$ and
$\displaystyle\frac{1}{n}\left[\vec{u}^{{}^{(\pm)}}_n
\cdot(\vec{D}^{F}\pm\vec{D}^{N})\right]^2$
at $L_F=L$, $L_N=0$ with different values of
$c/b$ for $\sigma_{\text{db}} = 0.6\%$,
and we observe that the case of $b\simeq c$ is well approximated
by the limit $b\ll c$.
 Therefore, from Eq.~(\ref{bllc}) we have
\begin{eqnarray}
\left(\begin{array}{c}
\tilde{\Lambda}_-\\
\tilde{\Lambda}_+
\end{array}\right)\simeq
\left(\begin{array}{c}
\displaystyle b\left[1-
\frac{(\vec{\tilde{v}}\cdot\vec{\tilde{h}})^2}{\vec{\tilde{v}}^2}\right]\\
\displaystyle c\,\vec{\tilde{v}}^2
\end{array}\right)
\simeq\left(\begin{array}{c}
\displaystyle b\left(1-
\frac{3.6^2}{24}\right)\\
\displaystyle 24c
\end{array}\right)
=\left(\begin{array}{c}
\displaystyle 0.46b\\
\displaystyle 24c
\end{array}\right).
\nonumber
\end{eqnarray}
 With the notations used in the text, we have
\begin{eqnarray}
\Lambda^{{}^{(-)}}_-/n&\simeq&\sigma^2_{\text{dB}}
\left[1-(\vec{v}\cdot\vec{h})^2/n\vec{v}^2\right]
\simeq0.46\sigma^2_{\text{dB}},\nonumber\\
\Lambda^{{}^{(-)}}_+/n&\simeq&\sigma^2_{\text{cal}}\vec{v}^2/n
\simeq24\sigma^2_{\text{cal}}.
\nonumber
\end{eqnarray}

\begin{acknowledgments}
The authors would like to thank Jacques Bouchez and
Anne Stutz for correspondence on the Bugey experiment.
This work was supported in part by Grants-in-Aid for Scientific Research
No.\ 16540260 and No.\ 16340078, Japan Ministry
of Education, Culture, Sports, Science, and Technology,
and the Research Fellowship
of Japan Society for the Promotion of Science
for young scientists.
\end{acknowledgments}


\newpage
\hglue -1.8cm
\renewcommand{\arraystretch}{2.5}
\begin{table}
\vglue 4.5cm
\hglue -1.8cm
\begin{tabular}{|l|c|c|c|}
\hline
 & optimized $L_F$/km & optimized $L_N$/km
& optimized value\\
\hline
$\displaystyle\frac{1}{n}
\sum_{j=1}^{n-2}\left[\vec{u}^{{}^{(-)}}_j\cdot(\vec{D}^{F}-\vec{D}^{N})\right]^2$
& 10.6 & 8.4 & 0.37 \\
$\displaystyle\frac{1}{n}
\left[\vec{u}^{{}^{(-)}}_{n-1}\cdot(\vec{D}^{F}-\vec{D}^{N})\right]^2$
& 2.1 & 0 & 0.41 \\
$\displaystyle\frac{1}{n}
\left[\vec{u}^{{}^{(-)}}_n\cdot(\vec{D}^{F}-\vec{D}^{N})\right]^2$
& 3.5 & 0 & 0.51 \\
$\displaystyle\frac{1}{n}
\sum_{j=1}^{n-2}\left[\vec{u}^{{}^{(+)}}_j\cdot(\vec{D}^{F}+\vec{D}^{N})\right]^2$
& 11.5 & 11.5 & 0.44 \\
$\displaystyle\frac{1}{n}
\left[\vec{u}^{{}^{(+)}}_{n-1}\cdot(\vec{D}^{F}+\vec{D}^{N})\right]^2$
& 2.1 & 2.1 & 1.6\\
$\displaystyle\frac{1}{n}
\left[\vec{u}^{{}^{(+)}}_n\cdot(\vec{D}^{F}+\vec{D}^{N})\right]^2$
& 3.5 & 3.5 & 2.1 \\
\hline
\end{tabular}
\vglue 1.63cm
\vglue 2.5cm
\caption{The values of each factor in the $\chi^2$ Eq.~(\ref{chi6})
in the spectrum analysis.  The number of bins is taken to be
$n=16$ here, but the values of these factors are almost independent
of $n$ for $n\gtrsim$8.  The energy interval adopted here
is 2.8MeV$\le E_\nu\le$7.8MeV,
as in the case of the Bugey experiment.\cite{Declais:1994su}}
\label{table}
\end{table}

\newpage

\newpage

\newpage

\newpage

\newpage

\end{document}